\documentclass[showpacs,amssymb,twocolumn,aps,nofootinbib]{revtex4}
\usepackage{amsmath}
\usepackage{amstext}
\usepackage{amsopn}
\usepackage{amsfonts}
\usepackage{amssymb}
\usepackage{bbm}
\usepackage{accents}
\usepackage{empheq}
\usepackage{graphicx}
\usepackage{epsf}
\usepackage{graphics}
\usepackage[latin1]{inputenc}

\newcommand{\be}{\begin{equation}}
\newcommand{\ee}{\end{equation}}
\newcommand{\bn}{\begin{eqnarray}}
\newcommand{\en}{\end{eqnarray}}
\newcommand{\bes}{\begin{subequations}}
\newcommand{\ees}{\end{subequations}}

\begin{document}

\title{A path integral approach to the Langevin equation}
\author{Ashok K. Das$^{a,b}$, Sudhakar Panda$^{c}$\footnote{on leave from HRI, Allahabad 211019, India.} and J. R. L. Santos$^{d,e}$}

\affiliation{$^a$ Department of Physics and Astronomy, University of Rochester, Rochester, NY 14627-0171, USA}
\affiliation{$^b$ Saha Institute of Nuclear Physics, 1/AF Bidhannagar, Calcutta 700064, India}
\affiliation{$^{c}$ Institute of Physics, Sachivalaya Marg, Bhubaneswar 751005, India}
\affiliation{$^d$Departamento de F\'{\i}sica, Universidade Federal da Para\'{\i}ba, 58051-970 Jo\~ao Pessoa, PB, Brazil}
\affiliation{$^e$Departamento de F\'{\i}sica, Universidade de Federal de Campina Grande, Para\'{\i}ba, Brazil}

\begin{abstract}
We study the Langevin equation with both a white noise and a colored noise. We construct the Lagrangian as well as the Hamiltonian for the generalized Langevin equation which leads naturally to a path integral description from first principles. This derivation clarifies the meaning of the additional fields introduced by Martin, Siggia and Rose in their functional formalism. We show that the transition amplitude, in this case, is the generating functional for correlation functions. We work out explicitly the correlation functions for the Markovian process of the Brownian motion of a free particle as well as for that of the non-Markovian process of the Brownian motion of a harmonic oscillator (Uhlenbeck-Ornstein model). The path integral description also leads to a simple derivation of the Fokker-Planck equation for the generalized Langevin equation.

\end{abstract}


\maketitle

\section{Introduction}
\label{sec_1}

The one dimensional Langevin equation, given by (a dot denotes a derivative with respect to time $t$) 
\begin{equation}
\dot{x} = v,\quad \dot{v} = - \frac{\partial S(v)}{\partial v} + \frac{\eta}{m},\label{1}
\end{equation}
describes a dynamical system subject to a random force (or noise) $\eta$ and $m$ denotes the mass of the particle \cite{langevin,uo,zwanzig,salinas,hj}. As a result of the presence of the random force, equation \eqref{1} becomes a stochastic differential equation. Namely, the dynamical variables can no longer be determined uniquely, rather they can be obtained as an average over the ensemble of the random noise. The simplest of the random forces is assumed to have a Gaussian probability distribution given by
\begin{equation}
P (\eta) = e^{-\frac{1}{4B} \int dt\, \eta^{2}(t)},\quad B > 0,\label{2}
\end{equation} 
so that all the odd correlation functions vanish and the even ones are completely determined by the two point correlation which has the form
\begin{equation}
\langle \eta (t) \eta (t')\rangle = 2 B \delta (t-t').\label{3}
\end{equation}
Such a noise is conventionally known as the white noise (Gaussian noise) and describes a Markovian process (a memoryless process). The simplest example of such a process is the Brownian motion of a free particle \cite{rbrown,einstein} described by \eqref{1}-\eqref{3} with $S(v) = \frac{1}{2} \gamma v^{2}$ where $\gamma >0$ corresponds to the coefficient of friction of the medium. One can also have more complicated random forces which are not given by a Gaussian distribution \cite{uo, zwanzig,hj, ms}. Such random forces are known as  colored (non-Gaussian) noise and in this case the two point correlation becomes a nontrivial function of time coordinates. Dynamical processes with colored noise are known as non-Markovian processes or processes with memory.

A simple example of a Langevin equation describing a non-Markovian process is obtained by generalizing \eqref{1} to
\begin{equation}
\dot{x} = v,\quad \dot{v} = - \frac{\partial S(v)}{\partial v} - \frac{1}{m}\frac{\partial V(x)}{\partial x} + \frac{\eta}{m},\label{4}
\end{equation}
with the random noise $\eta$ chosen to be Gaussian (see \eqref{2}-\eqref{3}). This system reduces to \eqref{1} for $V(x)=0$ (or a constant). The Brownian motion of a harmonic oscillator (also known as the Uhlenbeck-Ornstein  system \cite{uo}) is described by \eqref{4} with $S(v) = \frac{1}{2}\gamma v^{2}, \gamma >0$ and $V(x) = \frac{1}{2} m\omega^{2}x^{2}$ with $\omega$ representing the natural frequency of the oscillator. In this case the $v$ equation can be integrated to give
\begin{equation}
v (t) = \int_{-\infty}^{t} ds\, e^{-\gamma(t-s)} \left(-\omega^{2} x(s) + \frac{\eta(s)}{m}\right),\label{5}
\end{equation}
where we have chosen the initial time to be at infinite past. This, in turn, leads to an equation for $x$ of the form
\begin{equation}
\dot{x} = v = - \omega^{2} \int_{-\infty}^{t} ds\, e^{-\gamma(t-s)} x(s) + \frac{\overline{\eta}}{m},\label{6}
\end{equation}
where we have identified
\begin{equation}
\overline{\eta}(t) = \int_{-\infty}^{t} ds\, e^{-\gamma(t-s)} \eta(s).\label{7}
\end{equation}
Namely, the $x$ equation is described by a Langevin equation with an effective noise $\overline{\eta}$ given in \eqref{7} \cite{zwanzig,hj,ms}. Using \eqref{2},\eqref{3} and \eqref{7}, we can now show that
\begin{equation}
\langle \overline{\eta}(t) \overline{\eta} (t')\rangle = \frac{B}{\gamma}\,e^{-\gamma |t-t'|} = K(t-t'),\label{8}
\end{equation}
which is no longer Gaussian (even though the two point correlation of $\eta$ is, compare, for example, with  \eqref{3}).

It is worth noting here that the Langevin equation, which led to the development of the subject of stochastic differential equations, plays a central role in various investigations in different areas of physics including quantum field theory \cite{moffat}-\cite{gautier} and string theory \cite{son, rangamani}. The stochastic quantization proposed by Parisi and Wu \cite{parisi, nakano, bender, gozzi} is also based on a Langevin equation and provides an alternative quantization method which is particularly useful in the study of gauge theories. Here the idea is to study a quantum field theory as the equilibrium limit of a hypothetical stochastic process with respect to a fictitious time.  We note that  \cite{ckw} contains several examples of systems where the Langevin equation manifests and a very recent application in the context of the coupling of a heavy quark with a thermal anisotropic Yang-Mills plasma appears in \cite{chakrabortty}.

Generally, the correlation functions for the dynamical variables satisfying a Langevin equation (for example, in the case of the Brownian motion) are determined by first solving the dynamical equation in the presence of the random force and then averaging the product of the dynamical variables over the ensemble of the random noise \cite{uo,zwanzig,salinas}. We know that in the path integral description of a quantum theory the transition amplitude (generating functional) can be thought of as the  generator of correlation functions. Therefore, it is of interest to ask if a path integral description can be given to the Langevin equation so that one does not have to calculate individual correlation functions. Rather they can all be obtained directly from the generating functional (transition amplitude) by coupling to appropriate sources (as in quantum field theories). This is the question that we would like to study systematically in this paper and develop a path integral description for the Langevin equation from first principles. There is a second reason behind looking for a path integral description because one can then describe such a system (with a colored noise) in the closed time path formalism of Schwinger \cite{schwinger, das0} allowing us to (possibly) describe a general nonequilibrium quantum field theory (at finite temperature) and not just systems in the quenched approximation as have been mostly done so far.

Let us note here that a lot of work has been done in this direction starting with the work of \cite{martin}. The goal in \cite{martin} (as the authors themselves say) was not to derive a generating functional from first principles, rather to device a practical calculational method following from the well established formalisms of quantum field theory. In this approach, the functional equations for a stochastic process take the form of Schwinger-Dyson equations which do not define a closed system of equations. To make them a closed set, Martin, Siggia and Rose \cite{martin} introduced an additional pair of fields (operators) which do not commute with the original dynamical (stochastic) variables of the theory. For example, if $\psi$ denotes the doublet of dynamical variables $(x,v)$ satisfying the dynamical equation (explicit forms of $U_{2}, U_{3}$ depend on the specific model under study, we refer the readers to \cite{martin} for notations)
\begin{align}
\lefteqn{\dot{\psi}_{\alpha}(t) - \int dt' U_{2,\alpha\beta} (t,t') \psi_{\beta}(t')}\notag\\
& - \int dt' dt''\,U_{3,\alpha\beta\gamma} (t,t',t'')\psi_{\beta}(t')\psi_{\gamma}(t'') = U_{1,\alpha}(t),\label{8a}
\end{align}
then, \cite{martin} introduces an additional doublet of fields $\hat{\psi}(t)$ satisfying the quantization condition ($\hbar=1$ and $\alpha,\beta,\gamma = 1,2$)
\begin{equation}
[\psi_{\alpha} (t), \hat{\psi}_{\beta}(t)] =  \delta_{\alpha\beta},\label{8b}
\end{equation}
as well as the dynamical equation
\begin{align}
\lefteqn{\dot{\hat{\psi}}_{\alpha}(t) + \int dt' U_{2,\beta\alpha}(t',t)\hat{\psi}_{\beta}(t')}\notag\\
& + 2\int dt' dt''\, U_{3,\beta\gamma\alpha} (t', t'', t) \hat{\psi}_{\beta}(t') \psi_{\gamma}(t'') = 0.\label{8c}
\end{align}

However, although this makes the calculations of the correlation functions manageable, the physical meaning of the new operators as well as the equations is not very clear and leads to some unexpected behavior such as static properties depending on dynamical parameters of the theory. This formalism was further improved in \cite{dominicis} by combining with the ideas of renormalization group methods of quantum field theory which separate out the dependence of static quantities, but the meaning of the additional operators remains obscure. In some discussions (see, for example, \cite{zee}), one avoids the additional fields at the expense of adding nonlinearities into the theory which does not help much in the calculations. In \cite{bhattacharjee}, the issue of nonlinearities was handled in a different manner by using the method of stochastic quantization, but this method also has its own difficulties.

In this paper, we take a different approach, namely, we develop the path integral description of the Langevin equation starting from first principles \cite{das}. The first step in such a derivation is to determine the Lagrangian and the Hamiltonian for the system. In section {\bf II}, we describe this for the general Langevin system in \eqref{4} which makes clear that the additional fields (operators) and equations introduced in \cite{martin} coincide  exactly with those associated with the canonically conjugate variables of the starting dynamical degrees of freedom. We develop the path integral representation for the system in section {\bf III} where  we  discuss how the correlation functions can be obtained directly from the path integral both in the case of a Markovian process as well as a non-Markovian process. This is done explicitly for the cases of the Brownian motion of a free particle as well as for the case of the Brownian motion of a harmonic oscillator (Uhlenbeck-Ornstein system). In section {\bf IV}, we describe how a simple derivation of the Fokker-Planck equation for the general Langevin system can be obtained from the path integral representation of the system. We present a brief conclusion with future directions in section {\bf V}.

\section{The Lagrangian and the Hamiltonian}
\label{sec_2}
In  this section we will construct the Lagrangian and the Hamiltonian for the general Langevin equation given in \eqref{4}. This is the first step in developing a path integral description of the system \cite{das}. Let us recall that the general Langevin equation is given by
\be
\dot{x}=v\,;\qquad \dot{v}=-\frac{\partial\,S}{\partial\,v}-\frac{1}{m}\,\frac{\partial\,V}{\partial\,x}+\frac{\eta}{m},\label{2_1}
\ee
where $S=S(v(t)), V=V(x(t))$ and $\eta(t)$ is the noise function. It is clear that the dynamical equations in \eqref{2_1} can be obtained as the Euler-Lagrange equations (for $\xi$ and $\lambda$) from the Lagrangian
\be
{L}=\lambda\left(\dot{v}+\frac{\partial S}{\partial v}+\frac{1}{m}\frac{\partial V}{\partial x}-\frac{\eta}{m}\right)+\xi \left(\dot{x}-v\right),\label{2_2}
\ee
where $x, v, \xi$ and $\lambda$ are independent dynamical variables. The canonical conjugate momenta are obtained from the Lagrangian \eqref{2_2} to be 
\bn
&&
\Pi_x=\frac{\partial {L}}{\partial \dot{x}}=\xi,\qquad \Pi_v=\frac{\partial {L}}{\partial \dot{v}}=\lambda,\nonumber \\
&&
\Pi_\xi=\frac{\partial{L}}{\partial \dot{\xi}}=0,\qquad \Pi_\lambda=\frac{\partial{L}}{\partial \dot{\lambda}}=0. \label{2_3}
\en
Therefore, we have a constrained system with a set of four primary constraints
\begin{align}
\varphi_1 & = \Pi_x-\xi\simeq 0,\quad &  \varphi_2 & = \Pi_v-\lambda \simeq 0,\notag\\
\varphi_3 & = \Pi_\xi\simeq 0, \quad &  \varphi_{4} & = \Pi_\lambda \simeq 0.\label{2_4}
\end{align}

Consequently the canonical Hamiltonian has the form
\begin{align}
H_{can} & = \Pi_x\,\dot{x}+\Pi_v\,\dot{v}-{L} \nonumber\\
& = 
-\lambda\,\left(\frac{\partial\,S}{\partial\,v}+\frac{1}{m}\,\frac{\partial\,V}{\partial\,x}-\frac{\eta}{m}\right)+\,\xi\,v\,, \label{2_5}
\end{align}
and by adding the primary constraints \eqref{2_4} to $H_{can}$ we obtain the Hamiltonian for the system to be
\begin{align}
H & = H_{can}+\alpha_1\,\varphi_1+\alpha_2\,\varphi_2+\alpha_3\,\varphi_3+\alpha_4\,\varphi_4  \nonumber\\
& = 
-\lambda \left(\frac{\partial S}{\partial v}+\frac{1}{m}\,\frac{\partial V}{\partial x}-\frac{\eta}{m}\right)+ \xi v \nonumber\\
&\quad + 
\alpha_1 \left(\Pi_x-\xi\right) +\alpha_2 \left(\Pi_v-\lambda\right)+\alpha_3 \Pi_\xi+\alpha_4 \Pi_\lambda. \label{2_6}
\end{align}
Here $\alpha_{1},\alpha_{2},\alpha_{3},\alpha_{4}$ are Lagrange multipliers for the four primary constraints.

Requiring the primary constraints to be stationary does not lead to any secondary constraints, rather determines the four Lagrange multipliers, namely, 
\begin{align}
& 
\dot{\varphi_1}=\left\{\varphi_1,H\right\}=\frac{\lambda}{m}\,\frac{\partial^{2} V}{\partial x^{2}}-\alpha_3\simeq 0,\nonumber\\
&
\dot{\varphi_2}=\left\{\varphi_2,H\right\}=\lambda\,\frac{\partial^{2} S}{\partial v^{2}}-\xi-\alpha_4\simeq 0,\nonumber \\ 
&
\dot{\varphi_3}=\left\{\varphi_3,H\right\}=-v+\alpha_1 \simeq 0, \nonumber\\
& 
\dot{\varphi_4}=\left\{\varphi_4,H\right\}=\frac{\partial S}{\partial v}+\frac{1}{m}\frac{\partial V}{\partial x}-\frac{\eta}{m}+\alpha_2 \simeq 0. \label{2_7}
\end{align}
Substituting these back into \eqref{2_6}, the effective Hamiltonian takes the form
\begin{align}
H & = \Pi_\lambda \left(\lambda\,\frac{\partial^{2} S}{\partial v^{2}}-\xi\right)-\Pi_v \left(\frac{\partial S}{\partial v}+\frac{1}{m}\,\frac{\partial V}{\partial x}-\frac{\eta}{m}\right)\nonumber\\ 
& \quad +
\Pi_x v+\Pi_{\xi} \left(\frac{\lambda}{m}\,\frac{\partial^{2}\,V}{\partial x^{\,2}}\right).\label{2_8} 
\end{align}

The four primary constraints in \eqref{2_4} can be easily checked to be second class with 
\begin{align}
&
\left\{\varphi_1,\varphi_{3}\right\}=-1=-\left\{\varphi_{3},\varphi_{1}\right\},\nonumber\\
&
\left\{\varphi_2,\varphi_{4}\right\}=-1=-\left\{\varphi_{4},\varphi_{2}\right\}, \label{2_9}
\end{align}
so that the matrix of Poisson brackets takes the form ($a,b=1,2,3,4$)
\be
\left\{\varphi_a,\varphi_{b}\right\}
=C_{ab}=\left(\begin{array}{c c c c}
0 &\,\,\,\,\,0 & 		  -1  & \,\,\,\,\,0 \\
0 &\,\,\,\,\,0 &  \,\,\,\,\,0 &			 -1 \\
1 &\,\,\,\,\,0 &  \,\,\,\,\,0 & \,\,\,\,\,0 \\
0 &\,\,\,\,\,1 &  \,\,\,\,\,0 & \,\,\,\,\,0 \\
\end{array}
\right),\label{2_10}
\ee
with the inverse given by
\be 
C_{ab}^{-1}=\left(\begin{array}{c c c c}
\,\,\,\,\,0  &\,\,\,\,\,0& \,\,\,\,\,1 & \,\,\,\,\,0 \\
\,\,\,\,\,0  &\,\,\,\,\,0& \,\,\,\,\,0 & \,\,\,\,\,1 \\
-1           &\,\,\,\,\,0& \,\,\,\,\,0 & \,\,\,\,\,0 \\
\,\,\,\,\,0  &		  -1 & \,\,\,\,\,0 & \,\,\,\,\,0 \\
\end{array}
\right).\label{2_11}
\ee

The Dirac brackets between the dynamical variables can now be calculated from the definition
\be
\left\{A,B\right\}_D=\left\{A,B\right\}-\left\{A,\varphi_a\right\}C_{ab}^{-1}\left\{\varphi_{\,b},B\right\},\label{2_12}
\ee
and leads to
\begin{align}
&
\left\{x,\xi\right\}_D =1 = \left\{x,\Pi_{x}\right\}_{D},\nonumber\\
&
\left\{v,\lambda\right\}_D = 1 = \left\{v, \Pi_{v}\right\}_{D}, \label{2_13}
\end{align}
with all other brackets remaining unchanged. If we use the Dirac brackets, we can set all the constraints strongly equal to zero, which leads to the Hamiltonian
\be
H=-\lambda \left(\frac{\partial S}{\partial v}+\frac{1}{m}\,\frac{\partial V}{\partial x}-\frac{\eta}{m}\right)+\xi v.\label{2_14}
\ee
We note that the Hamiltonian in \eqref{2_14} indeed leads to the correct dynamical equations with the use of the Dirac brackets \eqref{2_13},
\begin{align}
\dot{x} & = \left\{x,H\right\}_D=\left\{x,\xi\right\}_D v=v,\notag\\
\dot{v} & = \left\{v,H\right\}_D=-\left\{v,\lambda\right\}_D\left(\frac{\partial S}{\partial v}+\frac{1}{m}\,\frac{\partial V}{\partial x}-\frac{\eta}{m}\right) \nonumber\\ 
& = 
-\frac{\partial S}{\partial v}-\frac{1}{m}\,\frac{\partial V}{\partial x}+\frac{\eta}{m}.\label{2_15}
\end{align}
Similarly, the dynamical equations for the conjugate variables $(\xi,\lambda)$ follow to correspond to
\begin{align}
\dot{\xi} & = \left\{\xi,H\right\}_D = \frac{\lambda}{m}\,\frac{\partial^{2}V(x)}{\partial x^{2}},\notag\\
\dot{\lambda} & = \left\{\lambda,H\right\}_D = -\xi + \lambda\,\frac{\partial^{2}S(v)}{\partial v^{2}}.\label{2_15a}
\end{align}

With these we are ready to make connection between the additional fields introduced in \cite{martin} (see eqs. \eqref{8a}-\eqref{8c}) and the conjugate variables associated with the dynamical variables $(x,v)$. First, we note that the Lagrangian in \eqref{2_2} as well as the associated Hamiltonian in \eqref{2_14} describe a first order system (much like a Dirac theory except that we are dealing with bosonic variables here) and if we identify $\psi$ with the doublet $(x,v)$ of dynamical variables and $\hat{\psi}$ with the corresponding doublet of the conjugate variables $(\xi, \lambda)$, the quantization condition in \eqref{2_13} can be written as
\begin{equation}
\left\{\psi_{\alpha} (t), \hat{\psi}_{\beta} (t)\right\}_D = \delta_{\alpha\beta},\label{2_15b}
\end{equation}
which coincides with \eqref{8b}. Furthermore, if we choose $S(v) = \frac{1}{2} \gamma v^{2},\ V(x) = \frac{1}{2} m\omega^{2}x^{2}$ corresponding to the damped harmonic oscillator and identify 
\begin{align}
& U_{1,\alpha} = \frac{\eta}{m} \delta_{\alpha 2},\notag\\ 
& U_{2,\alpha\beta}(t,t') = (\delta_{\alpha 1}\delta_{\beta 2} - \delta_{\alpha 2} (\omega^{2}\delta_{\beta 1}+\gamma\delta_{\beta 2}))\delta(t-t'),\notag\\
& U_{3,\alpha\beta\gamma} (t,t',t'') = 0,\label{2_15c}
\end{align}
equations \eqref{2_15} and \eqref{2_15a} can be written in the forms \eqref{8a} and \eqref{8c} respectively. On the other hand, if we choose $S(v)= \frac{1}{2} \gamma v^{2},\ V(x) = \frac{1}{2} m\omega^{2} x^{2} - \frac{1}{3} \nu x^{3}$ corresponding to a damped nonlinear oscillator and identify $U_{1}$ and $U_{2}$ as in \eqref{2_15c}, and
\begin{equation}
U_{3,\alpha\beta\gamma} (t,t',t'') = \frac{\nu}{m} \delta_{\alpha 2} \delta_{\beta 1} \delta_{\gamma 1} \delta(t-t')\delta(t-t''),\label{2_15d}
\end{equation}
then, eqs. \eqref{2_15} and \eqref{2_15a} take the forms of \eqref{8a} and \eqref{8c} respectively. This gives a natural physical meaning to the additional fields (operators) and equations introduced in \cite{martin}.

\section{Path integral description and correlations}
\label{sec_3}

Given the Hamiltonian and the Lagrangian, the transition amplitude in the presence of sources can be obtained in the standard manner \cite{das} to have the form
\be
U^{J}={\cal N}\int {\cal D}\eta\,{\cal D}x\,{\cal D}\xi\,{\cal D}v\,{\cal D}\lambda\,e^{iS^{J}-\frac{1}{4B}\int dt\,\eta^{2}},\label{3_1}
\ee
where we have denoted the sources generically by $J$ and have identified
\begin{equation}
S^{J} = \int dt\, L^{J},\label{3_2}
\end{equation}
with
\be
L^{J}=\lambda \left(\dot{v}+\frac{\partial S}{\partial v}+\frac{1}{m}\frac{\partial V}{\partial x}-\frac{\eta}{m}\right)+\xi\left(\dot{x}-v\right)+\widetilde{J} x+J v.\label{3_3}
\ee
Taking derivatives of the transition amplitude with respect to the sources $J$ and $\widetilde{J}$, we can calculate correlations involving $v$ and $x$ respectively. Here, in addition to the functional integration over the dynamical variables, we also have introduced a functional integral over the variable $\eta$ with a Gaussian weight to average over a white (Gaussian) noise (see \eqref{2}-\eqref{3}). It is understood that in calculating correlation functions the averaging over the noise is done at the end so that, in these calculations, the $\eta$ integrations has to be carried out only at the end. 

It is clear from \eqref{3_1} that the functional integration over $\lambda$ in the transition amplitudes leads to a delta function,
\begin{widetext}
\be
U^{J}={\cal N}\int {\cal D}\eta\,{\cal D}x\,{\cal D}\xi\,{\cal D}v\,\delta\left(\dot{v}+\frac{\partial S}{\partial v}+\frac{1}{m}\frac{\partial V}{\partial x}-\frac{\eta}{m}\right) e^{i\int dt \left[\xi\left(\dot{x}-v\right)+ \widetilde{J} x+ J v +\frac{i}{4B}\,\eta^{2}\right]}.\label{3_4}
\ee
\end{widetext}
Here, the delta function merely imposes the $v$-equation of motion in \eqref{4} which is not easy to solve in closed form for arbitrary nontrivial $S(v), V(x)$. Therefore, at this point we will demonstrate how the correlations are calculated from the path integral \eqref{3_1} by specializing to two examples. The first describes a Markovian process, namely, the Brownian motion of a free particle \eqref{1} while the second corresponds to a non-Markovian process, namely, the Uhlenbeck-Ornstein system or the Brownian motion of a harmonic oscillator \eqref{4} already noted in the introduction.

\subsection{Correlations for a Markovian process}

In this subsection, we will consider the Brownian motion of a free particle \cite{langevin, uo, hj, salinas} which corresponds to the choice 
\be
V(x)=0,\quad S(v)=\frac{1}{2}\,\gamma v^2,\quad \gamma > 0.\label{3_5}
\ee
In this case, the delta function constraint in the path integral \eqref{3_4} arising after the $\lambda$ integration implies
\be
\dot{v}+\gamma v-\frac{\eta}{m}=0,\label{3_6}
\ee
which has the analytical solution
\be
v_{c}(t)=v_0 e^{-\gamma t}+\frac{1}{m} \int_{0}^{\infty} dt^{\prime}\,G_{v} (t-t') \eta(t^{\prime}),\label{3_7}
\ee
where we have identified $t_i=0$ and $v_{c}(t_i)=v_{c}(0)=v_0$ and the first term in \eqref{3_7} represents the homogeneous solution. Furthermore, $G_{v}(t-t')$ denotes the (retarded) Green's function for \eqref{3_6} which has the explicit form
\begin{equation}
G_{v} (t-t') = \theta (t-t') e^{-\gamma (t-t')}.\label{3_8}
\end{equation}
Therefore, integrating over $v$ in the path integral using the delta function leads to the exponent in \eqref{3_4} of the form
\begin{equation}
i\int dt\left[\xi\left(\dot{x}-v_{c}\right)+ \widetilde{J} x+ J v_{c} +\frac{i}{4B}\,\eta^{2}\right],\label{3_9}
\end{equation}
and we note that the Jacobian arising from the delta function integration is trivial (identity).

Similarly, the integration over $\xi$ can be done trivially leading to the delta function constraint
\begin{equation}
\delta (\dot{x} - v_{c}),\label{3_10}
\end{equation}
which is solved by
\begin{equation}
x_{c} (t) = x_{0} + \left(\partial_{t}^{-1}v_{c}\right).\label{3_11}
\end{equation}
Here $x_{c} (t=0) = x_{0}$ and this allows us to do the integration over $x$ using the delta function to have the exponent of the form
\begin{equation}
i\int dt\left[\widetilde{J} x_{c} + J v_{c} + \frac{i}{4B}\, \eta^{2}\right].\label{3_12}
\end{equation}
Once again, the Jacobian coming from the delta function integration is trivial (identity). 

Substituting the values of $v_{c}$ and $x_{c}$ from \eqref{3_7} and \eqref{3_11} respectively (as well as $G_{v}$ from \eqref{3_8}), the transition amplitude takes the form
\begin{widetext}
\begin{equation}
U^{J} = 
{\cal N} e^{i\int dt\left[x_{0}\widetilde{J}+ v_0 \left(J-\left(\partial_t^{-1}\widetilde{J}\right)\right)e^{-\gamma t}\right]} \int {\cal D}\eta\,e^{-\frac{1}{4B}\int dt\,\eta^2 (t)+\frac{i}{m}\int dt\,dt^{\prime}\left(J-\partial_{t}^{-1}\widetilde{J}\right)\theta(t-t^{\prime})\,e^{-\gamma(t-t^{\prime})}\eta(t^{\prime})}.\label{3_13}
\end{equation}
\end{widetext}
The integrand is a Gaussian in $\eta$ (the exponent is at most quadratic in $\eta$) and, therefore, the functional integral over $\eta$ can be done leading to (the constant arising from the Gaussian integral has been absorbed into the normalization constant $\mathcal{N}$)
\begin{widetext}
\be	
U^{J}={\cal N} e^{i \int dt\left[x_{0}\widetilde{J} + v_0\left(J-\left(\partial_t^{-1}\widetilde{J}\right)\right) e^{-\gamma t}\right]} e^{-\frac{B}{2\gamma m^2}\int dt\,dt^{\prime} \left(J(t)-\partial_{t}^{-1}\widetilde{J}(t)\right)\left(J(t^{\prime})-\partial_{t^{\prime}}^{-1}\widetilde{J}(t^{\prime})\right)\left(e^{-\gamma\left|t-t^{\prime}\right|}-e^{-\gamma (t+t^{\prime})}\right)}.\label{3_14}
\ee
\end{widetext}

The transition amplitude $U^{J}$ can now be thought of as the generating functional for the correlation  functions of the theory involving $x$ and $v$. Let us start by determining the average value for $v(t)$, which can be obtained by taking the first derivative of $U^{J}$ with respect to $J(t)$ and then setting all the sources to zero \cite{das},
\be
\langle v(t) \rangle = \frac{(-i)}{U^{J}}\left. \frac{\delta U^{J}}{\delta J(t)}\right|_{\widetilde{J},J=0}=v_0\,e^{-\gamma t}.\label{3_15}
\ee
On the other hand, the second derivative of the generating functional with respect to $J(t)$ leads to 
\be
\langle v^2(t) \rangle =v_0^2\,e^{-2\gamma t}+\frac{B}{\gamma m^2} \left(1-e^{-2\gamma t}\right),\label{3_16}
\ee
as well as to (using \eqref{3_15})
\be
\langle \left(v(t)-\langle v(t)\rangle\right)^{2} \rangle=\frac{B}{\gamma m^2} \left(1-e^{-2\gamma t}\right).\label{3_17}
\ee
By taking higher order derivatives with respect to $J(t)$, one can directly obtain higher point correlation functions involving $v(t)$.

We note from \eqref{3_15}-\eqref{3_17} that, for $t\rightarrow \infty$, 
\be 
\langle v(t)\rangle = 0,\quad \langle v^2(t) \rangle =\langle \left(v(t)-\langle v(t)\rangle\right)^{2} \rangle=\frac{B}{\gamma m^2},\label{3_18}
\ee
namely, the average velocity vanishes for large time intervals while the mean square velocity approaches an equilibrium value \cite{uo, salinas}. From classical equipartition of energy in an equilibrium state at temperature $T$, we have (with the Boltzmann constant $k_B=1$)
\be
\frac{m}{2} \langle v^2(t) \rangle=\frac{T}{2},\label{3_19}
\ee
for a one dimensional ideal gas. Therefore, comparing with $\eqref{3_18}$ it follows that
\be 
B=\gamma m T,\label{3_20}
\ee
for large time intervals \cite{uo,zwanzig,einstein, salinas}.  

Let us next compute correlations involving the variable $x(t)$. By taking the derivative of the generating functional with respect to $\widetilde{J}(t)$ and setting all sources to zero, we obtain
\begin{align}
\langle x(t) \rangle & = \left.\frac{(-i)}{U^{J}}\frac{\delta U^{J}}{\delta \widetilde{J}(t)}\right|_{\widetilde{J},J=0} =
x_0+ v_{0}\left(\partial_t^{-1} e^{-\gamma t}\right)\notag\\ 
& = x_0 + \frac{v_0}{\gamma} \left(1-e^{-\gamma t}\right),\label{3_21}
\end{align}
where we have used $(\partial_{t}^{-1}f(t)) = \int_{0}^{t} dt'\,f(t')$. Similarly, by taking the second derivative with respect to $\widetilde{J}(t)$ (and setting all sources to zero), it straightforward to obtain
\begin{align}
\lefteqn{\langle x^2(t) \rangle = 
\left[x_0+\frac{v_0}{\gamma} \left(1-e^{-\gamma t}\right)\right]^2} \nonumber\\
& + 
\frac{B}{\gamma^2 m^2}\left[2t-\frac{2}{\gamma}\left(1-e^{-\gamma t}\right)-\frac{1}{\gamma}\left(1-e^{-\gamma t}\right)^2\right].\label{3_22}
\end{align}
We note here that, by taking higher order derivatives with respect to $\widetilde{J}$, we can obtain higher order correlation functions directly from the generating functional.

From \eqref{3_21} and \eqref{3_22}, we can now determine
\begin{align}
(\Delta x)^2 & = \langle (x(t)-\langle x(t) \rangle)^2 \rangle \nonumber\\
& = 
\frac{B}{\gamma^2 m^2} \left[2t-\frac{3}{\gamma} + \frac{4}{\gamma} e^{-\gamma t}-\frac{1}{\gamma} e^{-2\gamma t}\right],\label{3_23}
\end{align}
as well as
\begin{align}
\langle (x(t)- x_0 )^2 \rangle & = \frac{1}{\gamma^2} \left(v_0^2-\frac{B}{\gamma m^2}\right) \left(1-e^{-\gamma t}\right)^2 \notag\\ 
&\quad +
\frac{2B}{\gamma^2 m^2}\left[t-\frac{1}{\gamma}\left(1-e^{-\gamma t}\right)\right].\label{3_24}
\end{align}
Here \eqref{3_23} describes the mean square deviation of the coordinate $x$ while \eqref{3_24} gives the mean square displacement of the coordinate $x$ from its initial value $x_{0}$.

If the system is prepared in equilibrium (see \eqref{3_18}), we have
\be
v_0^2=\frac{B}{\gamma m^2},\label{3_25}
\ee
and \eqref{3_24} leads to
\be
\langle \left(x(t)-x_0\right)^2 \rangle=\left\{ \begin{array}{l c}
\frac{B}{\gamma m^2}\, t^{2},     & t \rightarrow 0, \\
							   &                  \\ 
\frac{2B}{\gamma^2 m^2}\, t, & t \rightarrow \infty. \\
\end{array}\right.\label{3_26}
\ee
Equations \eqref{3_23}, \eqref{3_24} and \eqref{3_26} imply that for $t\rightarrow \infty$ 
\be
\langle \left(x(t)-x_0\right)^2 \rangle=(\Delta x)^2=2Dt = \frac{2B}{\gamma^{2}m^{2}}\, t,\label{3_27}
\ee
where $D$ denotes the coefficient of diffusion and with \eqref{3_20} we recover Einstein's result
\be
D=\frac{B}{\gamma^2 m^2}=\frac{T}{\gamma m},\label{3_28}
\ee
which relates the coefficient of diffusion to the coefficient of friction (damping) in an inverse manner \cite{uo,zwanzig,einstein, salinas}. This relation can, in fact, be derived from the fluctuation-dissipation theorem in its classical form \cite{zwanzig,salinas,kubo,bettolo,kubo_c}.

\subsection{Correlations for a non-Markovian process}

As discussed in the introduction, the Uhlenbeck-Ornstein system \cite{uo,ms} or the Brownian motion of a harmonic oscillator provides a simple example of a non-Markovian process which we study in this subsection. Let us, therefore, choose (see discussion after \eqref{4})
\begin{equation}
S(v) = \frac{1}{2} \gamma v^{2},\quad V(x) = \frac{1}{2} m\omega^{2} x^{2},\label{3_29}
\end{equation}
where $\omega$ denotes the natural frequency of the oscillator. In this case, if we integrate over the $\xi$ variable in the path integral, it leads to the delta function constraint (see also \eqref{3_10})
\begin{equation}
\delta (\dot{x} - v),\label{3_30}
\end{equation}
which can be used to integrate out $v$. Next, the integration over $\lambda$ leads to the delta function constraint (see also \eqref{3_4})
\be 
\ddot{x}+ \gamma \dot{x}+\omega^2 x-\frac{\eta}{m}=0.\label{3_31}
\ee

This equation has two solutions - one a damped solution and the other a periodic damped solutions, where the period depends on $\gamma$ and $\omega$ in a special way. We choose to work with the periodic damped solution simply because this has been extensively discussed in \cite{uo} and we would like to compare our results to the ones derived there. (We note that the two solutions differ only in the homogeneous term.) The periodic damped solution has the form 
\begin{equation}
x_c(t) =  x_{h} (t) + \frac{1}{m} \int_0^{\infty} dt^{\prime} G_{x} (t-t') \eta(t^{\prime}),\label{3_32}
\end{equation}
where $x_{h}(t)$ represents the homogeneous solution of the form
\begin{equation}
x_{h} (t) = \left(x_0 \cos \omega_0 t + \frac{\gamma x_0+2v_0}{2\omega_0}\,\sin \omega_0 t\right)e^{-\frac{\gamma t}{2}},\label{3_33a}
\end{equation}
and $G_{x}$ denotes the (retarded) Green's function of \eqref{3_31}
\begin{equation}
G_{x} (t-t') = \theta (t-t')\,e^{-\frac{\gamma}{2}(t-t')}\,\frac{\sin \omega_{0}(t-t')}{\omega_{0}}.\label{3_33b}
\end{equation}
with
\be
\omega^{2} = \omega_0^{2} + \frac{\gamma^{2}}{4},\label{3_33c}
\ee
The integral over $x$ can now be done in the path integral and we note that the Jacobian coming from the delta function is trivial (unity).

Substituting all these into the transition amplitude, we obtain
\begin{widetext}
\begin{equation}
U^{J} = {\cal N} e^{i\int dt\,\left(\widetilde{J} + J\partial_{t}\right)x_{h} (t)}
 \int {\cal D}\eta\,e^{-\frac{1}{4B}\int dt\,\eta^{2}(t) + \frac{i}{m} \int dt\,dt'\left(\widetilde{J}(t)+ J(t)\partial_{t}\right) G_{x} (t-t') \eta(t')},\label{3_34}
\end{equation}
\end{widetext}
where $x_{h}(t)$ and $G_{x}(t-t')$ are defined in \eqref{3_33a} and \eqref{3_33b} respectively. We see that, as in the case of the  Brownian motion for the free particle (see \eqref{3_13}), the exponent is quadratic in $\eta$ and, therefore, the functional integral over $\eta$ can be done in a straightforward manner \cite{das}. The final result for the transition amplitude has the compact form
\begin{widetext}
\begin{equation}
U^{J} = {\cal N}\,e^{i\int dt\,(\widetilde{J}(t) + J(t)\partial_{t}) x_{h} (t) - \frac{B}{m^{2}}\int dt\,dt'\,dt''\left((\widetilde{J}(t') + J(t')\partial_{t'}) G_{x} (t' - t)\right)\left((\widetilde{J}(t") + J(t")\partial_{t"}) G_{x} (t" - t)\right)}.\label{3_35}
\end{equation}
\end{widetext}
Here the constant arising from the Gaussian integration has been absorbed into the normalization factor $\mathcal{N}$.

The transition amplitude in \eqref{3_35} can now be thought of as the generating functional for all the correlations in the theory. By taking the derivative with respect to $J(t)$ and setting all sources to zero, we can calculate the correlations involving $v(t)$. Thus, for example,
\begin{align}
\lefteqn{\langle v(t) \rangle = \left.\frac{(-i)}{U^{J}} \frac{\delta U^{J}}{\delta J(t)}\right|_{\widetilde{J}=J=0}  = (\partial_{t} x_{h} (t))}\nonumber\\	
& = 
\left(v_0 \cos(\omega_0 t)-\frac{2\omega^2 x_0+\gamma v_0}{2\omega_0} \sin(\omega_0 t)\right) e^{-\frac{\gamma t}{2}}.\label{3_36}
\end{align}
Similarly, by taking quadratic derivatives with respect to $J(t)$, we can also calculate the quadratic correlation for $v^2(t)$ which has the form
\begin{align}
\lefteqn{\langle v^2(t) \rangle = \left.\frac{(-i)^{2}}{U^{J}} \frac{\delta^{2}U^{J}}{\delta J(t) \delta J(t)}\right|_{\widetilde{J}=J=0}}\nonumber\\
& = \langle v(t)\rangle^2+\frac{B}{\gamma m^2} +\frac{B}{4\gamma m^2\,\omega_0^2}\nonumber\\
& \quad \times\left[\gamma^2 \cos(2\omega_0 t)+2\gamma \omega_0 \sin(2 \omega_0 t)-4\omega^2\right]e^{-\gamma t}.\label{3_37}
\end{align}
This shows that, as in the Markovian Langevin equation (see \eqref{3_18}), in the equilibrium regime ($t\rightarrow\infty$)
\be
\langle v(t)\rangle = 0,\quad \langle v^2(t) \rangle= \langle (v(t)-\langle v(t) \rangle)^2 \rangle = \frac{B}{\gamma m^2},\label{3_38}
\ee
which was also pointed out in \cite{uo}. Consequently, the constant $B$ has the same value as given in \eqref{3_20}. Any higher order correlation function involving only $v(t)$ can be obtained from the transition amplitude by taking higher order derivatives with respect to $J(t)$ and setting all the sources to zero. 

Let us next calculate the correlations involving the coordinate $x(t)$. By taking a derivative with respect to $\widetilde{J}(t)$ and setting all sources to zero we obtain
\begin{align}
\langle x(t) \rangle & = \left.\frac{(-i)}{U^{J}} \frac{\delta U^{J}}{\delta \widetilde{J}(t)}\right|_{\widetilde{J}=J=0} = x_{h}(t)\nonumber\\
& = 
\left[x_0 \cos(\omega_0 t)+\left(\frac{\gamma x_0+2v_0}{2\omega_0}\right)\sin(\omega_0 t)\right]e^{-\frac{\gamma t}{2}}.\label{3_39}
\end{align}
Similarly, the quadratic correlation $\langle x^2(t)\rangle$ is obtained by taking the second derivative of the transition amplitude with respect to $\widetilde{J}(t)$ and setting all sources to zero. This leads to
\begin{align}
\langle x^2(t) \rangle & =  \left.\frac{(-i)^{2}}{U^{J}} \frac{\delta^{2} U^{J}}{\delta \widetilde{J}(t) \delta \widetilde{J}(t)}\right|_{\widetilde{J}=J=0} \nonumber\\ 
& = 
\langle x(t) \rangle^2 + \frac{T}{m\omega^2}-\frac{T}{m\omega_0^{2}} e^{-\gamma t} \nonumber\\
& \quad + \frac{\gamma T}{4m\omega_0^{2}\omega^{2}}\left[\gamma \cos(2\omega_0 t)-2\omega_0\sin(2\omega_0 t)\right]e^{-\gamma t},\label{3_40}
\end{align}
where we used the value of $B$ as given in \eqref{3_20}. This is precisely the result obtained by Uhlenbeck and  Ornstein in \cite{uo}. In the equilibrium limit, it follows from \eqref{3_39} and \eqref{3_40} that 
\be
\langle x(t)\rangle = 0,\quad \langle x^2(t) \rangle = \langle (x(t)-\langle x(t) \rangle)^{2} \rangle=\frac{T}{m\omega^2},\label{3_41}
\ee
for the periodic damped solution of the generalized Langevin equation. By taking higher order derivatives of the transition amplitude with respect to $\widetilde{J}(t)$, one can similarly obtain higher correlation functions of $x(t)$.

The analysis in this section shows that the path integral representation of the transition amplitude in \eqref{3_1}-\eqref{3_3} can, in fact, be thought of as the generating functional for all the correlations of the generalized Langevin equation.

\section{The Fokker-Planck equation}
\label{sec_4}

The variables $x(t)$ and $v(t)$ in the Langevin equation \eqref{4} develop a probabilistic behavior because of the presence of the random noise $\eta (t)$ in the dynamical equation and one may be interested in the time evolution of such a probability distribution \cite{kubo_c}. Let $P(x,v;t)$ denote the probability for measuring the velocity for the Brownian particle to be $v$ and its position to be $x$ at time $t$. Then, for infinitesimally close time intervals, we can relate it to the transition amplitude as 
\be
P(x,v;t)={\cal N}\int dx^{\prime}\,dv^{\prime}\,U(x,v,t;x^{\prime},v^{\prime},t^{\prime})\,P(x^{\prime},v^{\prime},t^{\prime}),\label{4_1}
\ee
where $t^{\prime}$ is assumed to be infinitesimally earlier than $t$, namely,
\be
t^{\prime}=t-\epsilon,\qquad \epsilon\rightarrow 0^{+}.\label{4_2}
\ee

Let us next recall from \eqref{3_1} that the transition amplitude without sources is given by  
\be
U={\cal N}\int {\cal D}\eta\,{\cal D}x\,{\cal D}\xi\,{\cal D}v\,{\cal D}\lambda\,e^{i\int dt\,L - \frac{1}{4B}\int dt\,\eta^{2}}.\label{4_3}
\ee
Since we are not evaluating correlations (averages) here, we are not restricted to do the $\eta$ integration at the end. Furthermore, $\eta$ appears at most quadraically in the exponent so that it can be integrated out in a simple manner. Subsequently, the conjugate momenta $\lambda$ and $\xi$ can also be integrated out leading to
\be
U={\cal N}\int {\cal D}v\,{\cal D}x\,\delta\left(\dot{x}-v\right)e^{-\frac{m^2}{4B}\int dt \left(\dot{v}+\frac{\partial S}{\partial v}+\frac{1}{m}\frac{\partial V}{\partial x}\right)^2}.\label{4_4}
\ee
Therefore, the transition amplitude for an infinitesimal time interval is given by (there is no intermediate point to integrate over)
\begin{align}
\lefteqn{U = \delta\left(\frac{(x-x^{\prime})}{\epsilon} -v\right)}\nonumber\\
& \times e^{-\frac{m^2 \epsilon}{4B} \left[\left(\frac{v-v^{\prime}}{\epsilon}\right)^2
+2\left(\frac{\partial S}{\partial v'}+\frac{1}{m}\frac{\partial V}{\partial x'}\right)\frac{v-v^{\prime}}{\epsilon}
+\left(\frac{\partial S}{\partial v'}+\frac{1}{m}\frac{\partial V}{\partial x'}\right)^{2}\right]}. \label{4_5}
\end{align}
Here we are assuming normal ordering so that $V=V(x^{\prime})$ and $S=S(v^{\prime})$. (We note here that the Weyl ordering or the midpoint prescription leads to a slightly different, symmetric Fokker-Planck equation.) Substituting \eqref{4_5} into \eqref{4_1} and changing variables of integration as 
\be
x^{\prime}=x+\bar{\chi},\quad v^{\prime}=v+\chi,\label{4_6}
\ee
the probability distribution $P (x, v;t)$ takes the form
\begin{widetext}
\be
P(x,v;t)=\left(\frac{m^2}{4B\pi\epsilon^{3}}\right)^{\frac{1}{2}}\int d\bar{\chi} d\chi\, \delta\!\left(\frac{\bar{\chi}}{\epsilon}+v\right)e^{-\frac{m^2 \epsilon}{4B}\left[\left(\frac{-\chi}{\epsilon}\right)^2+2\left(\frac{\partial S(v+\chi)}{\partial v}+\frac{1}{m}\frac{\partial V(x+\bar{\chi})}{\partial x}\right)\frac{-\chi}{\epsilon}+\left(\frac{\partial S(v+\chi)}{\partial v}+\frac{1}{m}\frac{\partial V(x+\bar{\chi})}{\partial x}\right)^{2}\right]}P(x+\bar{\chi},v+\chi;t').\label{4_7}
\ee
\end{widetext}
where we have chosen the particular normalization constant for future use (basically it corresponds to the standard normalization for a single integration involving $v'$ as well as a factor of $\frac{1}{\epsilon}$ to offset the Jacobian $\epsilon$ arising from the $x'$ integration so that we have a nontrivial result).

It is clear from the delta function constraint as well as the first term in the exponent that
\begin{equation}
|\bar{\chi}| \sim \epsilon,\quad 0\leq |\chi| \leq \left(\frac{4B\epsilon}{m^{2}}\right)^{\frac{1}{2}}.\label{4_8}
\end{equation}
Therefore, if we want to calculate the infinitesimal change in the probability distribution (to obtain the time rate of change), it is clear that we can expand the integrand in powers of $\chi$ and $\bar{\chi}$ and keep terms only upto linear order in $\bar{\chi}$ and upto quadratic order in $\chi$ (without any $\epsilon$ factor multiplying). Keeping terms upto the relevant order, we obtain
\begin{widetext}
\begin{align}
P(x,v;t) & = 
\left(\frac{m^2}{4B\pi\epsilon^{3}}\right)^{\frac{1}{2}}\int d\bar{\chi}\,d\chi\,\delta\left(\frac{\bar{\chi}}{\epsilon} + v\right) e^{-\frac{m^2\chi^2}{4B\epsilon}}\left[1+\frac{m^2\chi}{2B}\frac{\partial S(v)}{\partial v}+\frac{m^2 \chi^2}{2B}\frac{\partial^{2} S}{\partial v^{2}}+\frac{m\chi}{2B} \frac{\partial V(x)}{\partial x}\right.\nonumber\\
& - 
\left. \frac{\epsilon}{4B}\left(\frac{\partial V(x)}{\partial x}\right)^{2}-\frac{m^2 \epsilon}{4B}\left(\frac{\partial S(v)}{\partial v}\right)^2+\left(\frac{m^4 \chi^2}{8\,B^{2}}\right)\left(\frac{\partial S(v)}{\partial v}\right)^{2}+\frac{m^{2}\chi^{2}}{8B^{2}}\left(\frac{\partial V(x)}{\partial x}\right)^{2} + O(\chi\bar{\chi})\right] \nonumber\\
& \times 
\left[ P(x,v;t^{\prime}) +\chi \frac{\partial P}{\partial v} +\frac{\chi^2}{2}\frac{\partial^{2} P}{\partial v^{2}}+\bar{\chi} \frac{\partial P}{\partial x}+O(\chi\bar{\chi})\right]. \label{4_9}
\end{align}
\end{widetext}

Now we  can do the integral over $\chi$ which involves a Gaussian as well as moments of the Gaussian. Similarly, the integral over $\bar{\chi}$ can be trivially done using the delta function constraint. So, keeping terms to linear order in $\epsilon$ we obtain
\begin{widetext}
\be
P(x,v;t)=P(x,v;t^{\prime})+\epsilon\left[-v \frac{\partial P}{\partial x}+\frac{B}{m^2} \frac{\partial^{2}P}{\partial v^{2}}+\frac{\partial}{\partial v} \left(\frac{\partial S(v)}{\partial v} P\right)+\frac{1}{m} \frac{\partial V(x)}{\partial x} \frac{\partial P}{\partial v}\right]+ O(\epsilon^{\frac{3}{2}}).\label{4_10}
\ee
\end{widetext}
This can be rewritten as 
\begin{align}
\lefteqn{\frac{P(x,v;t)-P(x,v;t^{\prime})}{\epsilon}=-v\frac{\partial P}{\partial x}+\frac{B}{m^2} \frac{\partial^{2}P}{\partial v^{2}}} \nonumber\\
& 
+\frac{\partial}{\partial v} \left(\frac{\partial S(v)}{\partial v} P\right)+\frac{1}{m} \frac{\partial V(x)}{\partial x} \frac{\partial P}{\partial v}+O(\epsilon^{1/2}), \label{4_11}
\end{align}
which, in the limit $\epsilon\rightarrow 0$, leads to the time evolution of the probability distribution
\be
\frac{\partial P}{\partial t}+v \frac{\partial P}{\partial x}-\frac{1}{m} \frac{\partial V}{\partial x} \frac{\partial P}{\partial v}=\frac{\partial}{\partial v} \left[\frac{B}{m^2} \frac{\partial P}{\partial v}+\frac{\partial S(v)}{\partial v} P\right].\label{4_12}
\ee
This indeed coincides with the Fokker-Planck equation for $P(x,v;t)$, as shown by Kubo \cite{kubo_c}.

\section{Conclusion}
\label{sec_5}

In this paper, we have developed the path integral description for the general Langevin equation starting from first principles. We have constructed the Lagrangian as well as the Hamiltonian for such a stochastic system and have constructed the path integral representation for the transition amplitude. This construction clarifies the physical meaning of the additional operators and equations introduced in \cite{martin} as those corresponding to the conjugate variables for the original dynamical variables of the system. We have shown, both for a Markovian process as well as a non-Markovian process, that the transition amplitude can be thought of as the generating functional for correlation functions. We have explicitly evaluated the correlations involving coordinate and velocity variables for the Brownian motion of a free particle as well as that for a harmonic oscillator (Uhlenbeck-Ornstein system). Furthermore, starting from the path integral description, we have also given a simple derivation of the Fokker-Planck equation for the time evolution of the probability distribution in the case of the general Langevin equation.

There remain some open questions which we would like to address in the future. For example, in this description (as is also the case in the standard treatments of these systems), the temperature dependence is brought into the problem through the fluctuation-dissipation theorem. So, in some sense the description is not yet quite complete. As a next step we would like to formulate the path integral using the closed time path formalism \cite{schwinger,das0} to see if the appropriate temperature dependence can be naturally generated. If not, it would be interesting to see how the fluctuation-dissipation theorem can be incorporated within this formalism so that it does not have to be imposed separately. If this description in the closed time path formalism works, it will open up the possibility of describing general nonequilibrium phenomena possibly through the use of colored noise. These are some of the questions that will be taken up in the future.

\acknowledgments

We thank the referee for bringing the earlier important works on the functional techniques to our attention. AD and JRLS would like to thank the Institute of Physics, Bhubaneswar and the University of Rochester respectively for hospitality where part of this work was done. JRLS was partly supported by CAPES and CNPq, Brazil.

\end{document}